\documentclass{emulateapj}

\begin{document}

\title{Discovery of an Unbound Hyper-Velocity Star in the Milky Way Halo}

\author{Warren R.\ Brown,
	Margaret J.\ Geller,
	Scott J.\ Kenyon, and 
	Michael J.\ Kurtz}

\affil{Smithsonian Astrophysical Observatory, 60 Garden St, Cambridge, MA 02138}
\email{wbrown@cfa.harvard.edu}

\slugcomment{Submitted to ApJ Letters}

\shorttitle{Discovery of an Unbound Hyper-Velocity Star}
\shortauthors{Brown et al.}


\begin{abstract}
	We have discovered a star, SDSS J090745.0+024507, leaving the
Galaxy with a heliocentric radial velocity of $+853\pm12$ km s$^{-1}$, the
largest velocity ever observed in the Milky Way halo.  The star is either
a hot blue horizontal branch star or a B9 main sequence star with a
heliocentric distance $\sim$55 kpc.  Corrected for the solar reflex motion
and to the local standard of rest, the Galactic rest-frame velocity is
$+709$ km s$^{-1}$.
	Because its radial velocity vector points 173.8$^{\circ}$ from the
Galactic center, we suggest that this star is the first example of a
hyper-velocity star ejected from the Galactic center as predicted by Hills
and later discussed by Yu \& Tremaine.  The star has [Fe/H]$\sim$0,
consistent with a Galactic center origin, and a travel time of
$\lesssim$80 Myr from the Galactic center, consistent with its stellar
lifetime.  If the star is indeed traveling from the Galactic center, it
should have a proper motion of 0.3 mas yr$^{-1}$ observable with {\it
GAIA}.  Identifying additional hyper-velocity stars throughout the halo
will constrain the production rate history of hyper-velocity stars at the
Galactic center.

\end{abstract}

\keywords{
        Galaxy: kinematics and dynamics---
        Galaxy: halo ---
        Galaxy: center ---
        Galaxy: stellar content ---
        stars: early type}

\section{INTRODUCTION}


	In a prescient paper, \citet{hills88} suggests that a stellar
binary interaction with the Milky Way's central black hole could eject one
member of the binary with a velocity $>1,000$ km s$^{-1}$.  \citet{yu03}
further develop Hill's analysis and suggest two additional mechanisms to
eject ``hyper-velocity'' stars from the Galactic center:  close
encounters of two single stars and three-body interactions between a
single star and a binary black hole.  \citet{yu03} predict production
rates for all three mechanisms.  Even the discovery of a single
hyper-velocity star can place important constraints on the formation
mechanism and the nature of the Galactic center.

	In our survey of faint blue horizontal branch candidates in the
Galactic halo, we have discovered a star, SDSS J090745.0+024507, traveling 
with a heliocentric radial velocity of $+853\pm12$
km s$^{-1}$.  Corrected to the local standard of rest and for the solar
reflex motion, the Galactic rest frame velocity of this star is
$v_{RF}=+709$ km s$^{-1}$.  The observed radial velocity is only a {\it
lower} limit to the star's true space velocity, but the radial velocity
alone substantially exceeds the escape velocity from the Galaxy.

	The distance to the hyper-velocity star (hereafter HVS) is $\sim$
55 kpc.  At a Galacto-centric distance of 50 kpc, the mass of the Milky
Way is 5.4$\times10^{11}$ M$_{\odot}$ \citep{wilkinson99} and the escape
velocity is 305 km s$^{-1}$.  Thus the HVS is moving well over {\it twice}
the escape velocity and in a direction 173.8$^{\circ}$ from the Galactic
center.

	By comparison, traditional ``high-velocity'' and ``run-away''
stars are stars with peculiar velocities greater than 30 km s$^{-1}$.  
High-velocity stars are typically early-type stars in the Galactic disk
moving away from star formation regions \citep[e.g.][]{hoogerwerf01}.  
Run-away B-type stars have been observed up to $\sim$15 kpc above the
Galactic plane and moving with radial velocities up to $\pm200$ km
s$^{-1}$
\citep{lynn04,magee01,ramspeck01,rolleston99,mitchell98,holmgren92,conlon90}.  
The highest velocity halo star, to our knowledge, was observed by
\citet{carney88} moving through the solar neighborhood with a total Galactic
rest frame velocity of 490 km s$^{-1}$.  In all cases, these high velocity
and run-away stars are very probably bound to the Galaxy.

	In \S 2 we describe the target selection and the spectroscopic
observations of the blue horizontal branch candidate sample.  The HVS is a
6$\sigma$ outlier from the distribution of radial velocities of this
sample.  In \S 3 we demonstrate the robustness of the radial velocity and
discuss the HVS's stellar properties.  In \S 4 we discuss the significance
of the star's hyper-velocity.

\section{TARGET SELECTION AND OBSERVATIONS}

	We have been using blue horizontal branch (BHB) stars to trace
velocity structure in the Milky Way halo \citep{brown04,brown03}.  In
2003, as part of an effort to measure the dynamical mass of the Milky Way
more accurately, we used Sloan Digital Sky Survey (SDSS) Early Data
Release and Data Release 1 photometry to select faint $19.75 < g'_0 <
20.25$ BHB candidates for spectroscopic observations.  We identified BHB
candidates by their A-type colors following \citet{yanny00}:  
$0.8<(u'-g')<1.5$, $-0.3<(g'-r')<0.0$.  Figure \ref{fig:ugr} shows this
color selection box, and colors of the 36 observed BHB candidates.  The
density of BHB candidates in this color/magnitude range is 0.3 objects
deg$^{-2}$.  Our observational strategy was to observe objects well-placed
in the sky at the time of observation.  Thus our sample of 36 stars is
``randomly'' selected from the area covered by the SDSS Early Data Release
and Data Release 1.

	We obtained spectra of the 36 BHB candidates with the 6.5m MMT
telescope during April, July, and December 2003.  We used the MMT Blue
Channel spectrograph with a 832 line/mm grating in second order.  This
set-up provides 1.0 \AA\ spectral resolution using a 1 arcsec slit.  With
a dispersion of 0.36 \AA/pix on the 2688$\times$512 CCD,

 \includegraphics[width=3.25in]{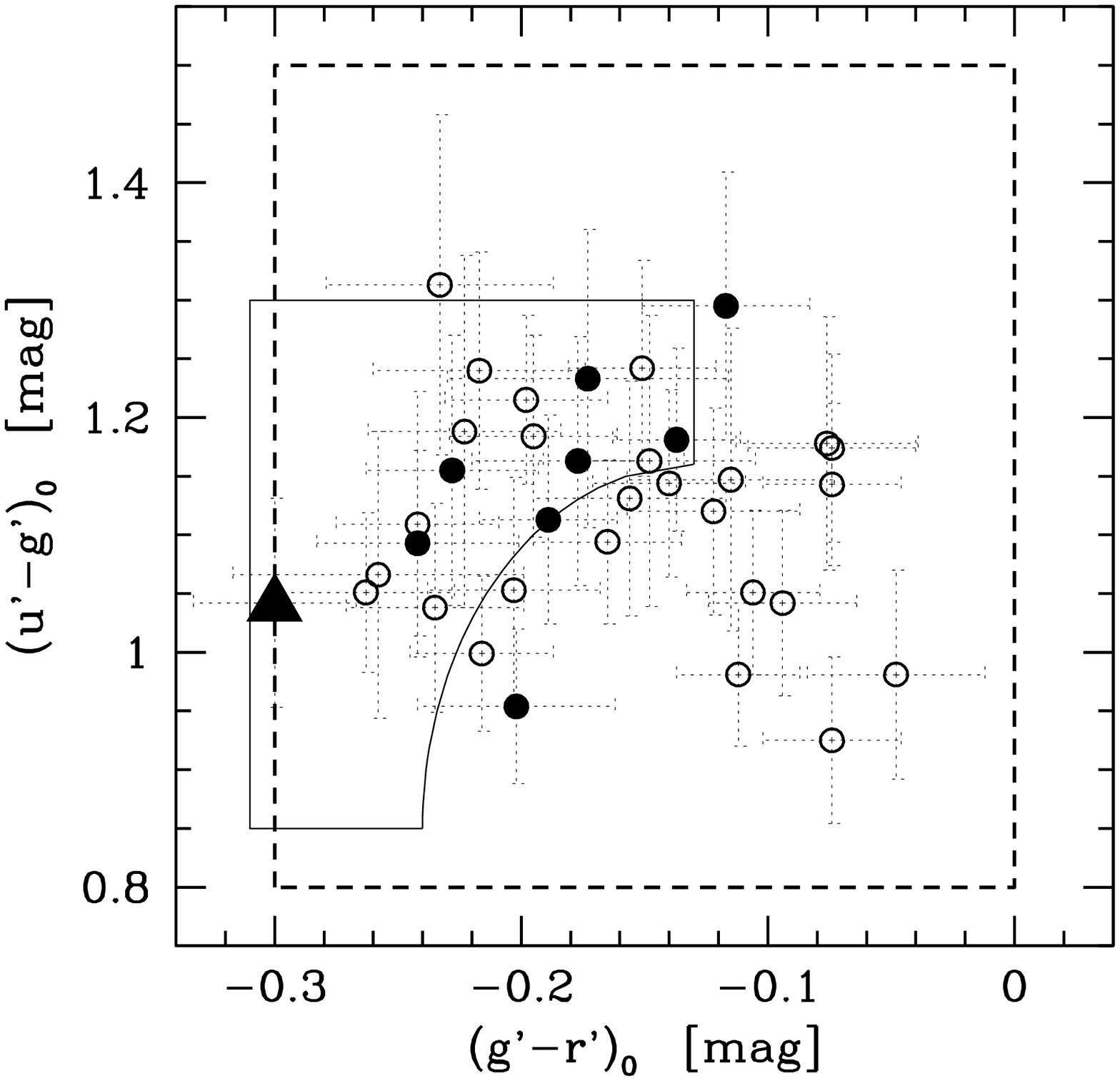}
 \figcaption{ \label{fig:ugr}
	Color-color plot of our sample.  The dashed box indicates
the selection box for A-type stars.  Points mark the colors of 
observed targets, with solid points marking the BHB stars.  Most of the 
solid points fall in the \citet{sirko04} BHB-selection box indicated by 
the solid line.  The solid triangle marks the HVS.}

	~

\noindent our spectral
coverage is 950 \AA.  All observations were made at the parallactic angle.  
On nights of good seeing $<1$ arcsec, we obtained $S/N \sim 20$ at 4,000
\AA\ in one hour of integration.

	The 36 BHB/A stars are all located in the Galactic halo with
heliocentric distances ranging $30<d<80$ kpc.  We measure heliocentric
radial velocities using the cross-correlation package RVSAO
\citep{kurtz98}.  We correct the velocities to the local standard of rest
\citep{dehnen98}.  We assume the stellar halo has no mean rotation, and
correct the velocities for the reflex motion of the local 220 km s$^{-1}$
orbital motion.

	Figure \ref{fig:veldisp} plots a histogram of the radial
velocities in the Galactic rest frame for all 36 BHB/A stars.  Ignoring
the HVS, the sample has a velocity dispersion of $\pm120$ km s$^{-1}$
consistent with a halo population.  The mean velocity of the sample
(ignoring the HVS) is $-7$ km s$^{-1}$, consistent with our assumption of
no rotation.  The $+709\pm12$ kms$^{-1}$ HVS, SDSS 090745.0+024507, is a
$6\sigma$ outlier from the observed distribution of radial velocities.

\section{THE HYPER-VELOCITY STAR}


	The HVS is located at $9^{h} 07^{m} 45\fs0,$ $+2^{\circ} 45\arcmin
07\arcsec$ (J2000).  In Galactic coordinates, the star is located at
$(l,b)=(227^{\circ} 20^{\arcmin} 07^{\arcsec}, +31^{\circ} 19\arcmin
55\arcsec)$.

	Fig \ref{fig:spectrum} shows the MMT Blue Channel spectrum of the
HVS.  The spectrum represents 60 minutes of integration
time, and has S/N=20 at 4,000 \AA.  The spectrum shows a raw heliocentric
radial velocity of $+853\pm12$ km s$^{-1}$.  Corrected to the local
standard of rest, the Galactic velocity components are $U=-491$ km
s$^{-1}$ (radially outwards), $V=-532$ km s$^{-1}$ (opposite the Galactic
rotation direction), and $W=+441$ km s$^{-1}$ (vertically upwards).

 \includegraphics[width=3.25in]{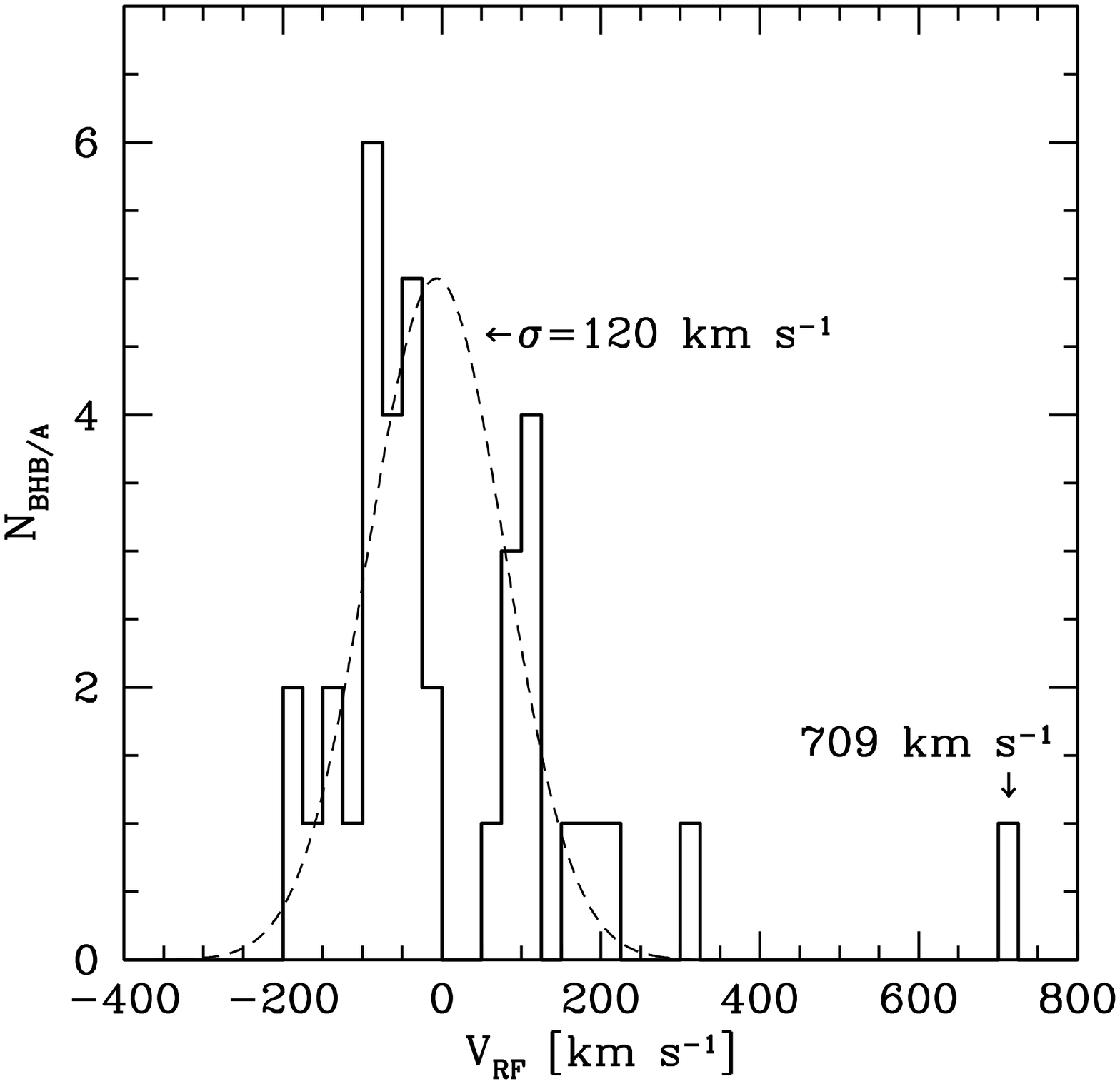}
 \figcaption{ \label{fig:veldisp}
	Galactic rest frame radial velocity distribution for our 36 BHB/A
star sample.  The HVS is a 6$\sigma$ outlier from the observed
distribution.}

	~

\subsection{Verifying the Radial Velocity}

	The substantial radial velocity is evident even from inspection of
the two-dimensional spectra.  Figure \ref{fig:twod} shows the $H\gamma$
line of the HVS shifted nearly on top of the night sky line at 4358.34
\AA.

	Moreover, the spectrum in Fig \ref{fig:spectrum} is not based on a
single observation, but on a series of three 20 minute observations obtained
over 1 hour.  The wavelength solution for the three observations is nearly
identical to the wavelength solutions obtained throughout the rest of that
night.  We verify the wavelength solution by measuring the velocity of the
night sky lines.  The Hg line at 4358.34 \AA\ (see Fig \ref{fig:twod}) has
the best S/N and a $-0.7 \pm 2.5$ km s$^{-1}$ velocity consistent with
zero.  We conclude that the wavelength solution is robust.

	We measure radial velocities for the three individual observations
and find that the velocities agree to $\pm5$ km s$^{-1}$.  Moreover, we
measure radial velocities using three different stellar templates (for B7,
B9, and A1 spectral types) and find that the velocities agree to $\pm3$ km
s$^{-1}$.  We conclude the radial velocity is accurate.  If the
HVS were a close binary, any systematic velocity offset is
likely to be small fraction of the observed velocity.

\subsection{Stellar Properties}

	The HVS has de-reddened colors $(u'-g')_0=1.04\pm0.09$ and
$(g'-r')_0=-0.30\pm0.03$, marked by the solid triangle in Fig
\ref{fig:ugr}.  These colors indicate it is probably a hot BHB star or a
late B-type main sequence star.  Its apparent magnitude is
$g'=19.81\pm0.02$.

	We classify the HVS's spectral type as B9.2 with an uncertainty of
1.2 spectral sub-types.  This classification is based on line indices
described in \citet{brown03}.  The spectral type is in perfect agreement
with the star's estimated effective temperature, $T_{eff} \sim 10,500$ K
(R.\ Wilhelm 2004, private communication).

	We measure the HVS's metallicity based on the equiv-

 \includegraphics[width=3.25in]{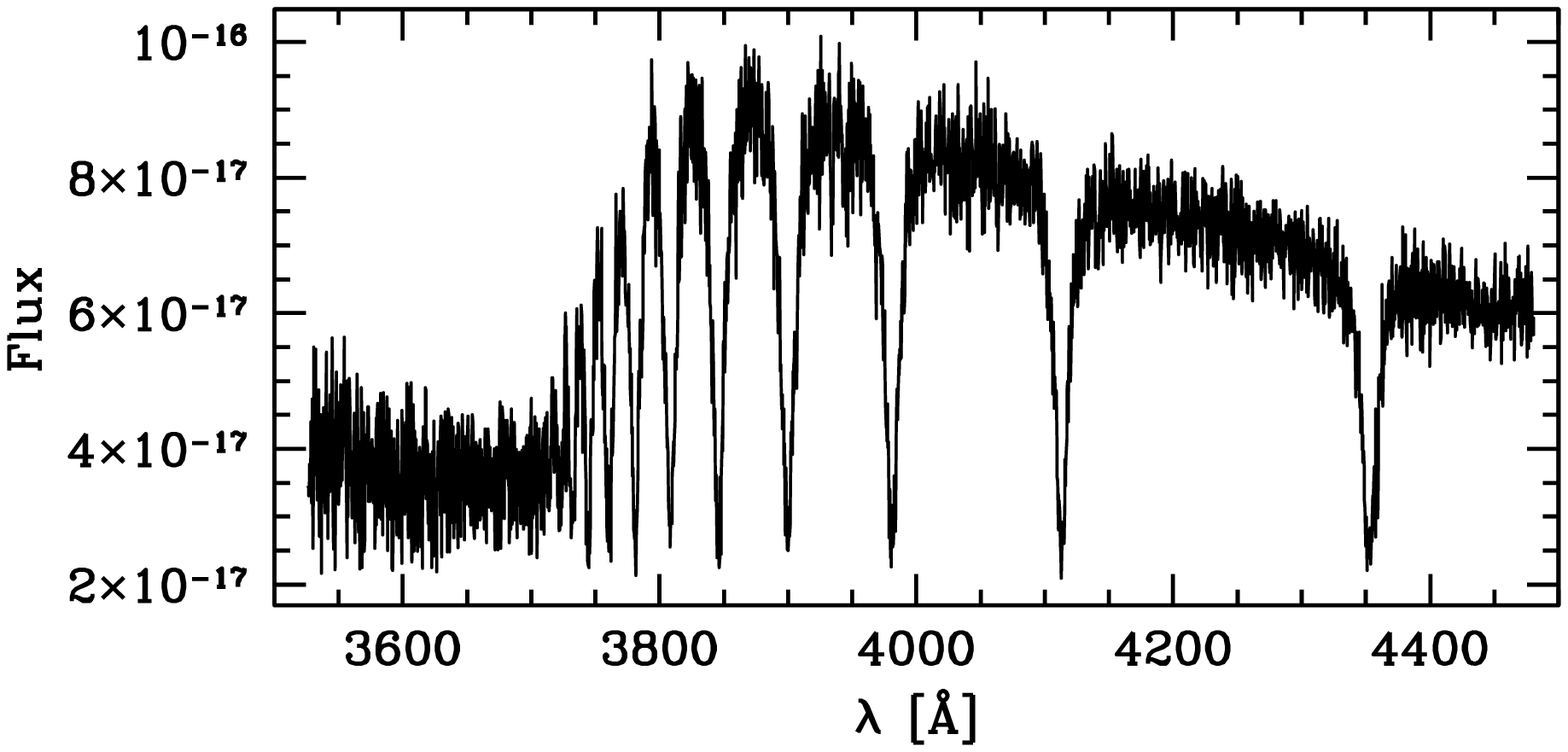}
 \figcaption{ \label{fig:spectrum}
	MMT spectrum of the HVS.}

	~

\noindent alent width of the Ca{\sc ii} K line and the star's photometric
colors.  However, the Ca{\sc ii} K line provides little leverage on [Fe/H]
at high $T_{eff}$. Thus there is considerable uncertainty in [Fe/H].  
Interpolated tables \citep{wilhelm99a} show that [Fe/H] can range from
-0.4 to well over 0.0; we conclude that the star's metallicity is roughly
solar [Fe/H]$\sim$0.0.  Solar metallicity suggests the HVS is probably a
B9 main sequence star.  However, given the uncertainty we cannot rule out
the star being a hot BHB star.

	We estimate the HVS's surface gravity by measuring the size and
steepness of its Balmer jump \citep{kinman94}, and the widths and the
shapes of its Balmer lines \citep{clewley02}.  These independent
techniques indicate that the star is a low surface gravity BHB star.  
However, classification by surface gravity is ambiguous at this $T_{eff}$.  
The $T_{eff}$ and $\log{g}$ of the horizontal branch crosses the main
sequence around $\sim$10,000 K.  Thus we cannot reliably distinguish
between a hot BHB or a B9 main sequence star based on its surface gravity.  
This uncertainty is problematic for estimating the star's distance:  a hot
BHB star and a B9 main sequence star with the same $T_{eff}$ and $\log{g}$
differ in luminosity by a factor of $\sim$4.

	We estimate the HVS's distance first assuming it is a hot BHB
star.  We calculate luminosity using the $M_V(BHB)$ relation of
\citet{clewley02}, which combines the \citet{gould98} {\it
Hipparcos}-derived $M_V$ zero point, the \citet{clementini03} LMC-derived
$M_V$-metallicity slope, and the \citet{preston91} $M_V$-temperature
correction.  If it is a hot BHB star, the HVS has $M_V(BHB)=1.9$ and a
heliocentric distance $d_{BHB}=39$ kpc.  Hot BHB stars are intrinsically
less-luminous stars that hook down off the classical horizontal branch. We
thus consider 39 kpc as a lower limit for the HVS's distance.

	We next estimate the HVS's distance assuming it is a B9 main
sequence star.  To estimate the luminosity of a B9 star we look at the
\citet{schaller92} stellar evolution tracks for a 3 $M_{\odot}$ star with
$Z=0.02$.  Such a star spends $3.5\times10^8$ yrs on the main sequence and
produces 160 $L_{\odot}$ when it has $T_{eff}\sim10,500$ K.  We convert
this luminosity to absolute magnitude using the bolometric corrections of
\citet{kenyon95}.  If it is a B9 main sequence star, the HVS has
$M_V(B9)=0.6$ and a heliocentric distance of $d_{B9}=71$ kpc.

	For purposes of discussion, we assume the HVS has the average
distance of these two estimates:  $d=55$ kpc.  This estimate places the
star at $z$ = $29 ~(d/55)$ kpc above the Galactic disk, and at $r$ = $60
~(d/55)$ kpc from the

 \includegraphics[width=3in]{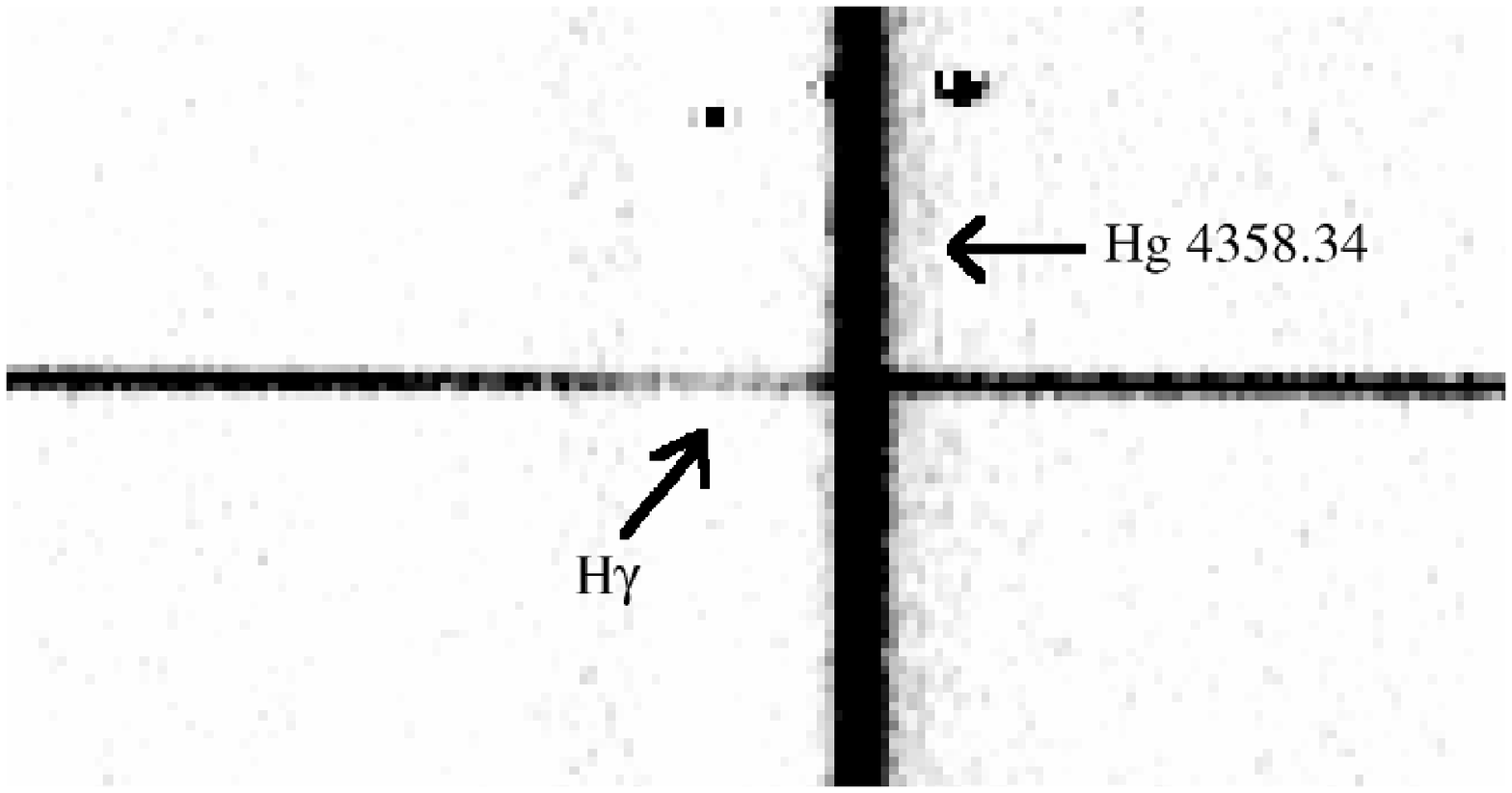}
 \figcaption{ \label{fig:twod}
	A portion of the two-dimensional spectrum.  The large radial 
velocity is evident from $H\gamma$ being shifted nearly on top of the 
night sky line Hg 4358.34 \AA.}

	~

\noindent Galactic center.

\section{DISCUSSION}

	Remarkably, \citet{hills88} predicted the existence of
hyper-velocity stars as a consequence of the presence of a massive black
hole at the Galactic center. Other mechanisms fail to produce velocities
as large as the one we observe for the HVS.  For example, the star is not
associated with the Sgr stream, nor is its radial velocity consistent with
any other member of the Local Group.  Association with high velocity
clouds is unlikely given their low velocity dispersion $\pm120$ km
s$^{-1}$ \citep{putnam02}.  The HVS is also unlikely to be in the tail of
the Galactic halo velocity distribution: our target selection yields
10$^4$ stars over the entire sky, and we should have to observe 10$^9$
objects to find one 6$\sigma$ outlier.  Supernova ejection
\citep{blaauw61} and dynamical ejection \citep{poveda67} are mechanisms
thought to produce run-away B stars, but both mechanisms predict maximum
ejection velocities of $\sim$300 km s$^{-1}$ \citep{leonard93}.

	\citet{hills88} predicts that tightly bound stellar binaries
encountering the central black hole can eject one star with velocity
$\gtrsim 1,000$ km s$^{-1}$ from the Galactic center.  \citet{yu03} also
predict that the close encounter of two single stars or the three-body
interaction between a single star and a binary black hole can result in
similar ejection velocities from the Galactic center.  \citet{yu03} show
that the production rates for these mechanisms are: (1) 10$^{-11}$
yr$^{-1}$ for single star encounters, (2) 10$^{-5}$ yr$^{-1}$ for binary
star encounters with the central black hole, and (3) 10$^{-4}$ yr$^{-1}$
for single star encounters with a binary black hole.  Thus the very
existence of the HVS rules out the single star encounter mechanism.

	Recent measurements of stellar orbits around the Galactic center
provide overwhelming evidence of a $4\times10^6$ M$_{\odot}$ black hole at
the Galactic center \citep{ghez05,schodel03,ghez03}.  Although it is
difficult to imagine a B star forming near the central black hole, many
young, massive stars {\it are} observed within 1 pc of the Galactic center
\citep[e.g.][]{genzel03}.  Moreover, one of the stars used to measure the
mass of the BH has a $45\pm16$ AU periapse and an orbital eccentricity 
$e=0.974\pm0.016$, giving it a periapse velocity $12,000\pm2,000$ km 
s$^{-1}$ \citep{ghez05}.

	The radial velocity vector of the HVS points 173.8$^{\circ}$ from
the Galactic center, suggestive of a Galactic center origin.  Even
assuming that the observed radial velocity is the full space motion of the
HVS, the travel time from the Galactic center is $\lesssim$80 Myr.  The
lifetime of a B9 main sequence star, by comparison, is approximately 350
Myr; the age of a star on the horizontal branch is much longer.  Thus the
star's solar metallicity, its direction of travel, and its travel time are
all consistent with a Galactic center origin.

	If the HVS is traveling on a radial path from the Galactic Center,
we predict its proper motion is $\sim 0.3 ~(d/55)$ mas yr$^{-1}$.  The
USNOB1 \citep{monet03} and the GSC2.3.1 (B.\ McLean, 2005 private
communication) catalogs list the star with a proper motion of $14\pm12$
and $20\pm11$ mas yr$^{-1}$, respectively, but in nearly opposite
directions.  The average of these proper motions is $3\pm8$ mas yr$^{-1}$
consistent with zero.  We searched for the star on 50 to 100 year-old
plates in the Harvard Plate Archive to increase the observed time
baseline, but the HVS was too faint.  The {\it GAIA} mission should be
able to observe 20 mag stars with 0.16 mas yr$^{-1}$ accuracy
\citep{perryman02} and thus may determine the full space motion of the
HVS.

	We can use our sample to place an upper limit on the production
rate of hyper-velocity stars.  Our sample of 36 BHB/A stars fills an
effective volume of $\sim$10$^3$ kpc$^3$ indicating an upper limit on the
density of hyper-velocity BHB/A stars, $\sim$10$^{-2}$ kpc$^{-3}$.  If it
takes 10$^8$ yr for a hyper-velocity star to leave the Galaxy, the
\citet{yu03} production rates imply a total of 10$^3$-10$^4$
hyper-velocity stars within the halo, a density of 10$^{-3}$-10$^{-2}$
kpc$^{-3}$.  At first glance there appears to be rough agreement between
the observed and predicted density of hyper-velocity stars.  However,
BHB/A stars represent only a small fraction of the total population of
halo stars, and there are essentially no constraints on the fraction of
the binary population they might represent near the Galactic center.

	The discovery of a HVS, as predicted by \citet{hills88} and
\citet{yu03}, provides an interesting piece of new evidence for a massive
black hole at the Galactic center.  Ironically, this evidence comes from
the radial velocity of a star $\sim$60 kpc from the Galactic center.  We
are now using the MMT Hectospec spectrograph, a 300 fiber spectrograph
with a 1$^{\circ}$ diameter field of view \citep{fabricant98hecto}, to
observe additional $g'_0\sim20$ BHB candidates over wide areas of sky.  
The identification of more hyper-velocity stars as a function of distance
and position on the sky will place better constraints on the production
mechanism and production rate of these unusual stars.  Observing
hyper-velocity stars over a wide range of distances will allow us to
measure the production history at the Galactic center and could probe the
history of the formation of the central black hole.

\acknowledgements
	We thank C.\ Heinke and A.\ Milone for their assistance with the
MMT observations.  We thank E.\ Falco for obtaining follow-up imaging at
the Whipple 1.2m telescope.  We thank T.\ Beers, D.\ Latham, B.\ McLean,
D.\ Mink, D.\ Monet, K.\ Stanek, and R.\ Wilhelm for helpful discussions.  
This work was supported by W.\ Brown's CfA Fellowship.


\begin{thebibliography}{37}
\expandafter\ifx\csname natexlab\endcsname\relax\def\natexlab#1{#1}\fi

\bibitem[{{Blaauw}(1961)}]{blaauw61}
{Blaauw}, A. 1961, \bain, 15, 265

\bibitem[{{Brown} {et~al.}(2003){Brown}, {Allende Prieto}, {Beers}, {Wilhelm},
  {Geller}, {Kenyon}, \& {Kurtz}}]{brown03}
{Brown}, W.~R., {Allende Prieto}, C., {Beers}, T.~C., {Wilhelm}, R., {Geller},
  M.~J., {Kenyon}, S.~J., \& {Kurtz}, M.~J. 2003, \aj, 126, 1362

\bibitem[{{Brown} {et~al.}(2004){Brown}, {Geller}, {Kenyon}, {Beers}, {Kurtz},
  \& {Roll}}]{brown04}
{Brown}, W.~R., {Geller}, M.~J., {Kenyon}, S.~J., {Beers}, T.~C., {Kurtz},
  M.~J., \& {Roll}, J.~B. 2004, \aj, 127, 1555

\bibitem[{{Carney} {et~al.}(1988){Carney}, {Laird}, \& {Latham}}]{carney88}
{Carney}, B.~W., {Laird}, J.~B., \& {Latham}, D.~W. 1988, \aj, 96, 560

\bibitem[{{Clementini} {et~al.}(2003){Clementini}, {Gratton}, {Bragaglia},
  {Carretta}, {Di Fabrizio}, \& {Maio}}]{clementini03}
{Clementini}, G., {Gratton}, R., {Bragaglia}, A., {Carretta}, E., {Di
  Fabrizio}, L., \& {Maio}, M. 2003, \aj, 125, 1309

\bibitem[{{Clewley} {et~al.}(2002){Clewley}, {Warren}, {Hewett}, {Norris},
  {Peterson}, \& {Evans}}]{clewley02}
{Clewley}, L., {Warren}, S.~J., {Hewett}, P.~C., {Norris}, J.~E., {Peterson},
  R.~C., \& {Evans}, N.~W. 2002, \mnras, 337, 87

\bibitem[{{Conlon} {et~al.}(1990){Conlon}, {Dufton}, {Keenan}, \&
  {Leonard}}]{conlon90}
{Conlon}, E.~S., {Dufton}, P.~L., {Keenan}, F.~P., \& {Leonard}, P.~J.~T. 1990,
  \aap, 236, 357

\bibitem[{{Dehnen} \& {Binney}(1998)}]{dehnen98}
{Dehnen}, W. \& {Binney}, J.~J. 1998, \mnras, 298, 387

\bibitem[{{Fabricant} {et~al.}(1998){Fabricant}, {Hertz}, {Szentgyorgyi},
  {Fata}, {Roll}, \& {Zajac}}]{fabricant98hecto}
{Fabricant}, D.~G., {Hertz}, E.~N., {Szentgyorgyi}, A.~H., {Fata}, R.~G.,
  {Roll}, J.~B., \& {Zajac}, J.~M. 1998, in Proc. SPIE 3355, Optical
  Astronomical Instrumentation, ed. S.~D'Odorico, 285--296

\bibitem[{{Genzel} {et~al.}(2003)}]{genzel03}
{Genzel}, R. {et~al.} 2003, \apj, 594, 812

\bibitem[{{Ghez} {et~al.}(2005){Ghez}, {Salim}, {Hornstein}, {Tanner}, {Lu},
  {Morris}, {Becklin}, \& {Duchene}}]{ghez05}
{Ghez}, A.~M., {Salim}, S., {Hornstein}, S.~D., {Tanner}, A., {Lu}, J.~R.,
  {Morris}, M., {Becklin}, E.~E., \& {Duchene}, G. 2005, \apj, accepted

\bibitem[{{Ghez} {et~al.}(2003)}]{ghez03}
{Ghez}, A.~M. {et~al.} 2003, \apjl, 586, L127

\bibitem[{{Gould} \& {Popowski}(1998)}]{gould98}
{Gould}, A. \& {Popowski}, P. 1998, \apj, 508, 844

\bibitem[{{Hills}(1988)}]{hills88}
{Hills}, J.~G. 1988, \nat, 331, 687

\bibitem[{{Holmgren} {et~al.}(1992){Holmgren}, {McCausland}, {Dufton},
  {Keenan}, \& {Kilkenny}}]{holmgren92}
{Holmgren}, D.~E., {McCausland}, R.~J.~H., {Dufton}, P.~L., {Keenan}, F.~P., \&
  {Kilkenny}, D. 1992, \mnras, 258, 521

\bibitem[{{Hoogerwerf} {et~al.}(2001){Hoogerwerf}, {de Bruijne}, \& {de
  Zeeuw}}]{hoogerwerf01}
{Hoogerwerf}, R., {de Bruijne}, J.~H.~J., \& {de Zeeuw}, P.~T. 2001, \aap, 365,
  49

\bibitem[{{Kenyon} \& {Hartmann}(1995)}]{kenyon95}
{Kenyon}, S.~J. \& {Hartmann}, L. 1995, \apjs, 101, 117

\bibitem[{{Kinman} {et~al.}(1994){Kinman}, {Suntzeff}, \& {Kraft}}]{kinman94}
{Kinman}, T.~D., {Suntzeff}, N.~B., \& {Kraft}, R.~P. 1994, \aj, 108, 1722

\bibitem[{{Kurtz} \& {Mink}(1998)}]{kurtz98}
{Kurtz}, M.~J. \& {Mink}, D.~J. 1998, \pasp, 110, 934

\bibitem[{{Leonard}(1993)}]{leonard93}
{Leonard}, P.~J.~T. 1993, in ASP Conf.\ Ser.\ 45, Luminous High-Latitude Stars,
  ed. D.~D. Sasselov, 360

\bibitem[{{Lynn} {et~al.}(2004){Lynn}, {Keenan}, {Dufton}, {Saffer},
  {Rolleston}, \& {Smoker}}]{lynn04}
{Lynn}, B.~B., {Keenan}, F.~P., {Dufton}, P.~L., {Saffer}, R.~A., {Rolleston},
  W.~R.~J., \& {Smoker}, J.~V. 2004, \mnras, 349, 821

\bibitem[{{Magee} {et~al.}(2001){Magee}, {Dufton}, {Keenan}, {Rolleston},
  {Kilkenny}, {O'Donoghue}, {Koen}, \& {Stobie}}]{magee01}
{Magee}, H.~R.~M., {Dufton}, P.~L., {Keenan}, F.~P., {Rolleston}, W.~R.~J.,
  {Kilkenny}, D., {O'Donoghue}, D., {Koen}, C., \& {Stobie}, R.~S. 2001,
  \mnras, 324, 747

\bibitem[{{Mitchell} {et~al.}(1998){Mitchell}, {Saffer}, {Howell}, \&
  {Brown}}]{mitchell98}
{Mitchell}, K.~J., {Saffer}, R.~A., {Howell}, S.~B., \& {Brown}, T.~M. 1998,
  \mnras, 295, 225

\bibitem[{{Monet} {et~al.}(2003)}]{monet03}
{Monet}, D.~G. {et~al.} 2003, \aj, 125, 984

\bibitem[{{Perryman}(2002)}]{perryman02}
{Perryman}, M.~A.~C. 2002, \apss, 280, 1

\bibitem[{{Poveda} {et~al.}(1967){Poveda}, {Ruiz}, \& {Allen}}]{poveda67}
{Poveda}, A., {Ruiz}, J., \& {Allen}, C. 1967, Bol.\ Obs\ Tonantzintla
  Tacubaya, 4, 860

\bibitem[{{Preston} {et~al.}(1991){Preston}, {Shectman}, \&
  {Beers}}]{preston91}
{Preston}, G.~W., {Shectman}, S.~A., \& {Beers}, T.~C. 1991, \apj, 375, 121

\bibitem[{{Putman} {et~al.}(2002)}]{putnam02}
{Putman}, M.~E. {et~al.} 2002, \aj, 123, 873

\bibitem[{{Ramspeck} {et~al.}(2001){Ramspeck}, {Heber}, \&
  {Moehler}}]{ramspeck01}
{Ramspeck}, M., {Heber}, U., \& {Moehler}, S. 2001, \aap, 378, 907

\bibitem[{{Rolleston} {et~al.}(1999){Rolleston}, {Hambly}, {Keenan}, {Dufton},
  \& {Saffer}}]{rolleston99}
{Rolleston}, W.~R.~J., {Hambly}, N.~C., {Keenan}, F.~P., {Dufton}, P.~L., \&
  {Saffer}, R.~A. 1999, \aap, 347, 69

\bibitem[{{Sch{\" o}del} {et~al.}(2003){Sch{\" o}del}, {Ott}, {Genzel},
  {Eckart}, {Mouawad}, \& {Alexander}}]{schodel03}
{Sch{\" o}del}, R., {Ott}, T., {Genzel}, R., {Eckart}, A., {Mouawad}, N., \&
  {Alexander}, T. 2003, \apj, 596, 1015

\bibitem[{{Schaller} {et~al.}(1992){Schaller}, {Schaerer}, {Meynet}, \&
  {Maeder}}]{schaller92}
{Schaller}, G., {Schaerer}, D., {Meynet}, G., \& {Maeder}, A. 1992, \aaps, 96,
  269

\bibitem[{{Sirko} {et~al.}(2004)}]{sirko04}
{Sirko}, E. {et~al.} 2004, \aj, 127, 899

\bibitem[{{Wilhelm} {et~al.}(1999){Wilhelm}, {Beers}, \& {Gray}}]{wilhelm99a}
{Wilhelm}, R., {Beers}, T.~C., \& {Gray}, R.~O. 1999, \aj, 117, 2308

\bibitem[{{Wilkinson} \& {Evans}(1999)}]{wilkinson99}
{Wilkinson}, M.~I. \& {Evans}, N.~W. 1999, \mnras, 310, 645

\bibitem[{{Yanny} {et~al.}(2000)}]{yanny00}
{Yanny}, B. {et~al.} 2000, \apj, 540, 825

\bibitem[{{Yu} \& {Tremaine}(2003)}]{yu03}
{Yu}, Q. \& {Tremaine}, S. 2003, \apj, 599, 1129

\end{thebibliography}


\end{document}